\documentclass[%
 reprint,
nofootinbib,
 amsmath,amssymb,
 aps,
]{revtex4-2}

\usepackage{graphicx}
\usepackage{dcolumn}
\usepackage{bm}

\begin{document}

\preprint{APS/123-QED}

\title{Constraints on the Equation of State of Nuclear Matter from Systematically Comparing SMASH Calculations to HADES Data}

\author{Justin Mohs$^1$}
\email{jmohs@itp.uni-frankfurt.de}
\author{Simon Spies$^2$}
\author{Hannah Elfner$^{3,1,4,5}$}
\email{h.elfner@gsi.de}

\address{1 Institute for Theoretical Physics, Goethe University, Max-von-Laue-Strasse 1, 60438 Frankfurt am Main, Germany}

\address{2 Institute for Nuclear Physics, Goethe University, Max-von-Laue-Strasse 1, 60438 Frankfurt am Main, Germany}

\address{3 GSI Helmholtzzentrum f\"ur Schwerionenforschung, Planckstr. 1, 64291 Darmstadt, Germany}
\address{4 Frankfurt Institute for Advanced Studies, Ruth-Moufang-Strasse 1, 60438 Frankfurt am Main, Germany}
\address{5 Helmholtz Research Academy Hesse for FAIR (HFHF), GSI Helmholtz Center, Campus Frankfurt, Max-von-Laue-Straße 12, 60438 Frankfurt am Main, Germany}

\date{\today}

\begin{abstract}
We aim to constrain the equation of state of nuclear matter by comparing calculations with the SMASH transport model to directed and elliptic flow measurements for protons and deuterons performed by the HADES collaboration in a systematic way.
A momentum-dependent term is included in the potential for which we show that it is needed to describe flow data.
We further incorporate a simple symmetry potential in the transport model and present constraints on the stiffness of the equation of state of nuclear matter at saturation density and on the symmetry potential.
The constraints are obtained by performing a Bayesian analysis such that we can also provide an uncertainty for the estimated parameters.
The posterior distribution is obtained by Markov chain Monte Carlo sampling for which we emulate the transport model with a Gaussian process to lower computational costs.
We find that a relatively stiff equation of state is favoured in our analysis with a small uncertainty whereas the constraints obtained for the symmetry potential are rather loose.

\end{abstract}

\maketitle

\section{Introduction}
\label{sec:intro}
The equation of state (EoS) of nuclear matter is a research topic of great interest as it encodes fundamental properties and is a main ingredient for the description of the dynamics of nuclear matter.
In addition, the interest in the equation of state of nuclear matter has increased with the first observation of gravitational waves originating from a merger of neutron stars \cite{LIGOScientific:2017vwq}.
It was further shown that nuclear matter can be probed under similar conditions in heavy ion collisions at SIS18 energies as in neutron star mergers \cite{HADES:2019auv}, meaning that findings about the EoS from heavy-ion collisions can contribute to understanding astrophysical events.

Knowledge about the EoS can only be obtained using a model to relate the equation of state to observables of a heavy ion collision.
As they can be applied to systems out of equilibrium, transport models are commonly used for the modeling of heavy-ion collisions and therefore to put constraints on the equation of state of nuclear matter.

Since findings about the EoS come from different fields of physics, it is very valuable to estimate the uncertainty of a result such that it can be used for a combined analyses such as \cite{Tsang:2023vhh,Huth:2021bsp,Dutra:2012mb,Dutra:2014qga}.
Bayesian inference is an excellent method to evaluate the equation of state including uncertainties given experimental data.
Recent works provide constraints on the EoS using Bayesian inference \cite{OmanaKuttan:2022aml, Oliinychenko:2022uvy} but also simpler methods of estimating the uncertainty have proven to be extremely useful for combined analyses \cite{Danielewicz:2002pu,LeFevre:2015paj,Lynch:2009vc}.

Two main experimental measurements commonly used for extracting information about the EoS are strangeness production \cite{Hartnack:2005tr,Hartnack:2011cn,Fuchs:2000kp} and the transverse anisotropy of proton and light nuclei distributions \cite{Molitoris:1985gs,Aichelin:1987ti,Welke:1988zz,Hillmann:2018nmd,Hillmann:2019wlt,Mohs:2020awg,Nara:2021fuu,Steinheimer:2022gqb,Tarasovicova:2024isp}.
Not all of the constraints agree however, which might be caused by implementation choices and different ingredients in different models.
Therefore, much effort is put into comparing transport models within the transport model evaluation project \cite{TMEP:2022xjg,SpiRIT:2020sfn}.
It has also been shown that just the amount of resonances present in the model can have a strong influence on the extracted equation of state \cite{Hombach:1998wr}.

In this work, we aim to provide a constraint on the EoS by comparing calculations with the SMASH transport model to flow data on protons and deuterons from the HADES experiment.
A comparison was already done in a previous work \cite{Mohs:2020awg} but no satisfactory agreement with the measurement could be achieved and the conclusion was drawn that a momentum-dependent part in the nuclear potential is required.
Here, we present results including a momentum-dependence and found a good description of the experimental data.
As better agreement with the measurement can be achieved, we go forward to perform a Bayesian inference for the incompressibility of the EoS and the symmetry potential.
This study is restricted to experimental data from the HADES experiment but a study comparing to measurements from the FOPI experiment has been carried out recently \cite{Tarasovicova:2024isp}. 
The influence of the centrality selection on the observables is studied and a prescription that closely follows the one applied in the HADES experiment is compared with the centrality selection via impact parameter.

We start with a brief introduction of the model in Section \ref{sec:model_description}, where we focus on the nuclear potentials applied in this work.
Here, we also give details on the light nuclei formation, which is performed using coalescence in this work.
Next, different ingredients of the model and the comparison to data are investigated in Section \ref{sec:sensitivity} and a final choice for the ingredients of the analysis is made based on the findings.
Further, we define the ingredients of the Bayesian inference before show result in Section \ref{sec:results}.
Finally, we formulate conclusions in Section \ref{sec:conclusions}.

\section{Model Description}
\label{sec:model_description}
This work aims to put systematic constraints on the equation of state by comparing recent data from the HADES experiment with calculations performed with the SMASH transport approach version 3.1.
A detailed description of the model can be found in the main reference \cite{SMASH:2016zqf} and the source code is publicly available \cite{wergieluk_2024_10707746}.
In the following, we briefly address main features of the model and go into slightly more detail on the nuclear potentials as they are most relevant for the equation of state.

The degrees of freedom in SMASH are hadrons of which we adopt the properties from the particle data group \cite{ParticleDataGroup:2018ovx}.
Regarding the collision term, the most important interactions are elastic scatterings and resonance formation.
We include a large variety of resonances with over 200 species plus the different isospin states and their respective antiparticles.
A full list is publicly available in human-readable format with the code base \cite{wergieluk_2024_10707746}.
We apply vacuum Breit-Wigner spectral functions for all resonances with their width and pole mass adopted from PDG \cite{ParticleDataGroup:2018ovx} and fitted to elementary cross-section data.
For the interactions, Pauli-blocking is taken into account with more details to be found in the main SMASH reference \cite{SMASH:2016zqf}.

\subsection{Nuclear potentials in SMASH}
The nuclear potentials used in this work include a Skyrme and a Symmetry part of the form
\begin{equation}
    U_{\mathrm{Skyrme}} = A \left(\frac{\rho_B}{\rho_0} \right) + B\left( \frac{\rho_B}{\rho_0}\right)^\tau
    \label{eq:skyrme}
\end{equation}
\begin{equation}
    U_\mathrm{Sym} = \pm 2S_\mathrm{pot} \frac{\rho_{I3}}{\rho_0}\,.
    \label{eq:symmetry}
\end{equation}
Here, $\rho_B$ is the baryon density and the, $\rho_{I3}$ is the density of the relative isospin projection and the sign depends on the isospin of the particle of interest. $A$, $B$, $\tau$ and $S_\mathrm{pot}$ are parameters which can be related to the equation of state.
We further add a momentum-dependent term to our potential, for which we use the form as suggested by Welke et al. \cite{Welke:1988zz}
\begin{equation}
U_\mathrm{MD}(\rho,\mathbf{p})=\frac{2C}{\rho_0}g\int\frac{d^3p'}{(2\pi)^3}\frac{f(\mathbf{r}, \mathbf{p}')}{1+\left(\frac{\mathbf{p}-\mathbf{p}'}{\Lambda}\right)^2} \,.
\label{eq:momentum_dependence}
\end{equation}
The momentum-dependent term incorporates the constant parameters $C$ and $\Lambda$.
One further requires the distribution function $f(\mathbf{r},\mathbf{p})$ for which we follow the implementation in GiBUU \cite{Buss:2011mx} and assume cold nuclear matter $f(\mathbf{r},\mathbf{p}) = \Theta(p_F(\rho_B(\mathbf{r}))-|\mathbf{p}|)$ and, consistently, $g=4$ for the spin and isospin degeneracy of nucleons.
This simplification allows for an analytical solution of the integral (see \cite{Welke:1988zz}) which reduces the computation time to a reasonable amount.

The parameters $A$, $B$, $C$, $\tau$ and $\Lambda$ are required to fulfil the nuclear ground-state properties
\begin{equation}
    \left(\frac{E_b}{A}\right)_{\rho_B=\rho_0} = -16\,\rm MeV
\end{equation}
\begin{equation}
    \frac{\partial}{\partial \rho_B}\left(\frac{E_b}{A}\right)_{\rho_B=\rho_0}=0
\end{equation}
with the binding energy per nucleon $E_b/A$.
We further adopt the constraints
\begin{equation}
    U(\rho_0, p=0)=-75\,\mathrm{MeV} 
\end{equation}
\begin{equation}
    U(\rho_0,p=800\,\mathrm{MeV}) = 0\,\mathrm{MeV}
\end{equation}
taken from \cite{Welke:1988zz} that are chosen to match the optical potential, which we verified comparing the Schr\"oder equivalent potential to the analysis by Cooper et al. \cite{Cooper:2009zza}.
We then require a given incompressibility $\kappa=9\rho^2\frac{\partial^2}{\partial \rho^2}\left(\frac{E_b}{A}\right)_{\rho=\rho_0}$ as the final equations to fix the parameters.
In Section \ref{sec:results}, we use experimental data to fix the two remaining parameters of the potentials $\kappa$ and $S_\mathrm{pot}$.

Finally, we include a Coulomb potential in a relatively simple form by evaluating the Poisson equation in integral form and apply the Biot-Savart law by summing over electric charges and currents on a lattice, effectively assuming magnetostatics.

All potentials are expressed in terms of densities.
We therefore require a smooth density profile with sufficient statistics to perform a reliable calculation.
In order to achieve that, we apply a Gaussian smearing kernel in a covariant form as described in the appendix of \cite{Oliinychenko:2015lva}.
Sufficient statistics in the density calculation is achieved by using 1000 parallel ensembles while we represent each real particle by a single particle, which makes the application of coalescence for light nuclei more straightforward.

\subsection{Light nuclei formation}

Light nuclei are created from protons and neutrons by a coalescence requirement. This requirement is adjusted to match the proportional amounts of protons, deuterons, tritons, and helium-3 within the HADES acceptance. It is defined by a momentum difference of less than 300 MeV/c and a distance in coordinate space of less than 3 fm at the time of the last elastic or inelastic scattering. Additionally, a momentum penalty is introduced if a cluster includes two or more protons, in order to simulate the Coulomb repulsion.

The clustering procedure itself is performed similar to the minimum spanning tree approach: First, all deuteron candidates (proton neutron pairs that fulfill the coalescence requirement) are identified and the best one is selected. Next, cluster candidates involving the selected cluster and a further nucleon fulfilling the coalescence requirement that would create a cluster with a mean lifetime of at least 100 fm/c are identified and the best candidate is selected for the next iteration. These two step are repeated iteratively until no further cluster candidates are found. To ensure that only clusters which would traverse the particle detectors of an experimental setup are written to the output, final clusters are only accepted if they have a mean lifetime above 100 ns. Finally, the four-vector-masses of the created clusters are adjusted to the nominal masses of the cluster by assuming single photon emission.

\subsection{Parameter extraction}
The goal of this work is to find constraints on the EoS, here entering as parameters of the potentials, in a controlled way.
For this endeavor, we apply Bayes theorem to find a posterior distribution for the model parameters given the experimental data set and prior knowledge.

As mentioned in Section \ref{sec:intro}, some information on the equation of state is already present from previous work.
We aim, however, to provide a constraint from only HADES data that can in future work be combined with different constraints from other experiments and from other research fields.
Hence, we choose a prior that is constant in $\kappa$ and $S_\mathrm{pot}$ in a region that we consider plausible.
This region is defined by $200\,\mathrm{MeV} < \kappa < 400 \,\mathrm{MeV}$ and $0<S_\mathrm{pot} < 30\,\mathrm{MeV}$.
Softer incompressibilities have been reported before but in section \ref{sec:results} we will observe that a rather stiff equation of state is favoured and the lower end of the prior is not affecting the results in any way.
The range for $S_\mathrm{pot}$ was chosen to allow for a large range around the default value of $S_\mathrm{pot} = 18\,\mathrm{MeV}$ that was suggested for a code comparison \cite{TMEP:2016tup} and serves us as a starting point.

The other main ingredient for Bayes theorem is the likelihood which is the probability to observe the experimental measurement given a set of parameters.
We write the likelihood like the $\chi^2$ as typically used for fitting
\begin{equation}
    P(\mathrm{data}|\kappa,S_\mathrm{pot}) \propto \prod_i \frac{1}{\sqrt{2\pi\sigma_i^2}}\exp\left(-\frac{(y_i^\mathrm{model} - y_i^\mathrm{data})^2}{2\sigma_i^2}\right)\,.
\end{equation}
The index $i$ runs over all data points included in the analysis with corresponding experimental values $y_i^\mathrm{data}$ and model predictions $y_i$.
The uncertainty $\sigma_i$ is the sum of the squares of the experimental uncertainty (systematic + statistical) and the statistical model uncertainty $\sigma_i^2 = (\sigma_i^\mathrm{data})^2 + (\sigma_i^\mathrm{model})^2$.
The included observables are directed and elliptic flow of protons and deuterons for several rapidity-bins and $1\,\mathrm{GeV} < p_T < 1.5\,\mathrm{GeV}$ as published by the HADES collaboration \cite{HADES:2020lob}.
This observable was selected because the big transverse momentum bin allows for a calculation of the flow coefficients with a reasonably small number of events and larger momenta are more sensitive to the equation of state, whereas lower momenta are strongly affected by the treatment for light nuclei formation \cite{Mohs:2020awg}. In general, the nucleon flow is most sensitive to the nuclear potentials, while their yield is affected by light nuclei formation to a large extent. The deuteron flow allows to obtain insights on the symmetry energy. 

As the evaluation of the model is computationally very expensive, we create a Gaussian process emulator to interpolate training data from SMASH.
This structure is commonly used to perform Bayesian parameter estimation using computationally expensive models in the field \cite{Bernhard:2015hxa,Bernhard:2016tnd,JETSCAPE:2020mzn}.
Our analysis aims to only determine two free parameters, which allows us to train the Gaussian process with only few samples of training data.
We found that the emulator gives very reliable results for only 15 samples, which we distribute in the parameter space using Latin hypercube sampling.
The uncertainties of the predictions from the Gaussian process are used as $\sigma_i^\mathrm{model}$ in the likelihood function and we validate the predictions in Section \ref{sec:results}.

\section{Sensitivity Study}
\label{sec:sensitivity}
In this section, different features of the model and their influence on flow observables are investigated.
The aim is to make a well informed decision on technical options regarding the comparison to experimental data and to study the importance of including physical features, namely the Coulomb potential and the momentum-dependent term in the potential.

All model curves shown in this section are obtained by training a Gaussian process on real SMASH calculations and evaluating that Gaussian process.
A verification of the Gaussian process is presented for the final model choice in Section \ref{sec:results} but such a verification has been performed for curves presented here.
We start by investigating differences in the elliptic flow emerging from different methods of performing the centrality selection.

\subsection{Centrality selection}
The classification of events into centrality intervals is an important task since the directed and elliptic flow signals are induced by the geometry of the system.
In that sense, the impact parameter to a large extend determines the anisotropy in the final state.
We compare in this section two methods of performing the centrality selection.

The first one is by using a fixed range of impact parameters for each centrality class as it is extracted from the Glauber model.
The impact parameter ranges used for the comparison to gold-gold collisions at the HADES experiment have been determined in \cite{HADES:2017def}.
The centrality selection in the experiment is, however, performed event by event based on the number of hits in the detector.
Naturally, small impact parameters correlate to more activity in the detector but not every single event with a small impact parameter yields a larger number of tracks in the detector than an event with a larger impact parameter.
Clearly, it is worthwhile to investigate how large the difference is if the activity in the detector is simulated for every individual event of the model calculation and the events are grouped into centrality classes based on that.
For this purpose, the number of detector hits is calculated in a large number of minimum-bias events in the model.
The events are then grouped into centrality classes based on the number of hits and a mapping from the number of hits to the corresponding centrality class is created.
This way, the centrality selection can be performed based on detector hits even if the model does not necessarily describe the produced number of particles perfectly.

\begin{figure}
    \centering
    \includegraphics[width=\linewidth]{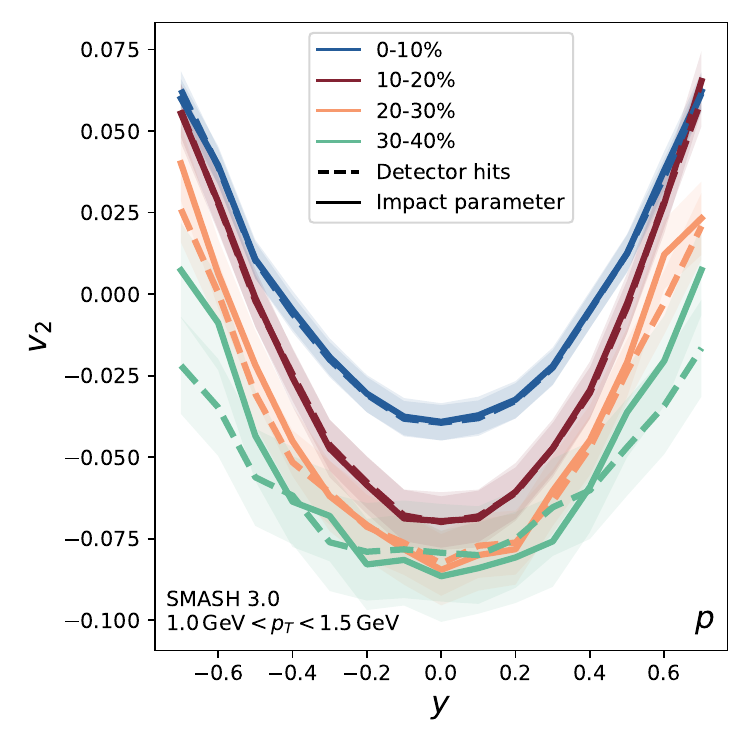}
    \caption{Directed flow of protons for different centrality intervals in gold-gold collisions at $E_\mathrm{kin}=1.23A\,\mathrm{GeV}$.
    Compared are different methods for the classification into centrality intervals.
    The full lines depict the centrality selection based on the impact parameter.
    It is compared to the dashed lines which represent the selection based on the simulated number of tracks in the detector.
    The curves are created based on a Gaussian emulator trained on SMASH calculations.
    The parameters of the potentials are for this figure chosen to be in the center of the prior range.}
    \label{fig:centrality_comparison}
\end{figure}

Taking the acceptance and efficiency of the detector into account, such a centrality determination is performed here.
A comparison between the centrality selection via the impact parameter and via the number of detector hits is shown in Figure \ref{fig:centrality_comparison} for the elliptic flow of protons in different centrality classes.
For this observable, mostly no difference is found for the elliptic flow except in the forward and backward region but even there the difference is statistically not significant.
Figure \ref{fig:centrality_comparison} does not include experimental data points because both calculations were performed without the momentum-dependent term that is essential for a reasonable description of the data as we show in Section \ref{sec:mom_dep_sensitivity}.
We perform this comparison without momentum-dependence because the mapping from detector hits to centrality classes has been generated for the potentials without momentum-dependence and we observe that the dynamics of the system change quite significantly when including the momentum-dependent part, such that the other mapping cannot be applied without modification.

As we do not observe a significant difference between the two methods in Figure \ref{fig:centrality_comparison}, we perform the centrality selection based on the impact parameter in the following calculations.
Still the method is very interesting as it is very close to the experiment and it may be important for different observables or other kinematic ranges. 

\subsection{Coulomb Potential}
\begin{figure}
    \centering
    \includegraphics[width=\linewidth]{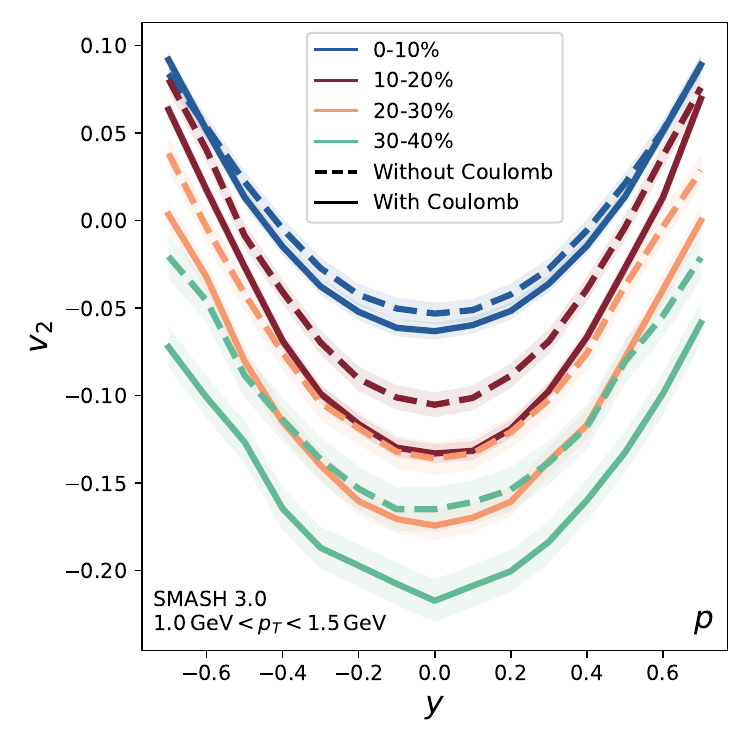}
    \caption{Elliptic flow of protons for different centrality intervals in gold-gold collisions at $E_\mathrm{kin}=1.23A\,\mathrm{GeV}$.
    The elliptic flow is compared between a calculation including the Coulomb potential and a calculation disregarding it.
    The curves are created based on a Gaussian emulator trained on SMASH calculations.
    The parameters of the potentials are for this figure chosen to be in the center of the prior range.}
    \label{fig:coulomb_comparison}
\end{figure}

Charged particles interact electromagnetically.
Clearly this should be taken into account for a realistic calculation but we want to study here how big the impact of electromagnetic potentials is.
The thought behind this sensitivity study is that the long range of the electromagnetic interaction makes is numerically quite expensive to include in a BUU-type calculation.
The difference between a calculation for the elliptic flow of protons including the Coulomb potential and a calculation neglecting the Coulomb potential is presented in Figure \ref{fig:coulomb_comparison}.
While the observed elliptic flow is still rather close in central events, a sizable effect is observed in other centrality classes.
As the system is positively charged, the Coulomb potential asserts a repulsive force, so that the elliptic flow signal is enhanced due to the inclusion of the Coulomb potential.

We conclude that the inclusion of the Coulomb potential is worth the numerical costs.
The results presented in Section \ref{sec:results} are obtained with the Coulomb potential.
This is for this study especially needed because the symmetry potential is treated as a parameter of the Bayesian analysis and only when the Coulomb potential is included in the calculation, one can cleanly extract the contribution of the symmetry energy.

\subsection{Momentum-dependent Potentials}
\label{sec:mom_dep_sensitivity}

The optical model has been used to extract constraints on the momentum-dependence of the nuclear potential at saturation density from $p+A$ scattering experiments in \cite{Hama:1990vr,Cooper:2009zza}.
The potential extracted this way can of course only be described including a momentum-dependence in the potentials.

We still want to investigate the influence of the momentum-dependent term on flow observables in the SMASH model.
In this section, we further aim to show how good of an agreement with experimental data can be achieved with and without momentum-dependence when the parameters are tuned to the experimental data.

\begin{figure}
    \centering
    \includegraphics[width=\linewidth]{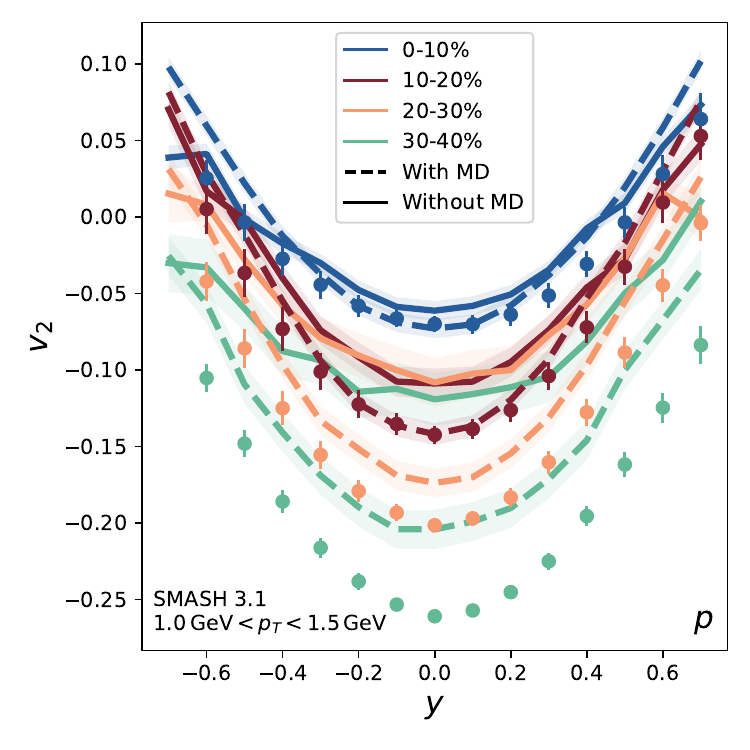}
    \caption{Elliptic flow of protons for different centrality bins in gold-gold collisions at $E_\mathrm{kin}=1.23A\,\mathrm{GeV}$.
    A calculation with momentum-dependence is compared to a calculation without the momentum-dependent term.
    All curves are predicted by a Gaussian process which was trained on SMASH output.
    For both models the parameters that give the best overall description of the data shown in Figure \ref{fig:observables} are chosen.}
    \label{fig:momentum_dependence_comparison}
\end{figure}

For this purpose, we perform the Bayesian analysis, that is described in more detail for the full model in Section \ref{sec:results}, with and without momentum-dependence.
We find the maximum of the posterior distribution which is the parameter set that has the most overlap with the experimental data.
In Figure \ref{fig:momentum_dependence_comparison} we present the elliptic flow of protons for the best possible parameter set including momentum-dependence and for the best parameter set without momentum-dependence.
Note that the parameters are not tuned only to only this observable but to $v_1$ and $v_2$ of protons and deuterons (see Fig. \ref{fig:observables}).
For the sensitivity test presented in Figure \ref{fig:momentum_dependence_comparison}, the Coulomb potential is neglected but one can clearly see that the data cannot be described without momentum-dependence in SMASH.
When the momentum-dependent part is included, one observes a stronger elliptic flow signal.
It is still weaker than the experimental measurement but a further improvement is presented in Section \ref{sec:results}, where the Coulomb potential is added.

All results presented in the following include the momentum-dependent part of the potential as it is necessary to be consistent with the findings from the optical model and strongly enhances the agreement with experimental data. 

\section{Results}
\label{sec:results}

\begin{figure}
    \centering
    \includegraphics[width=\columnwidth]{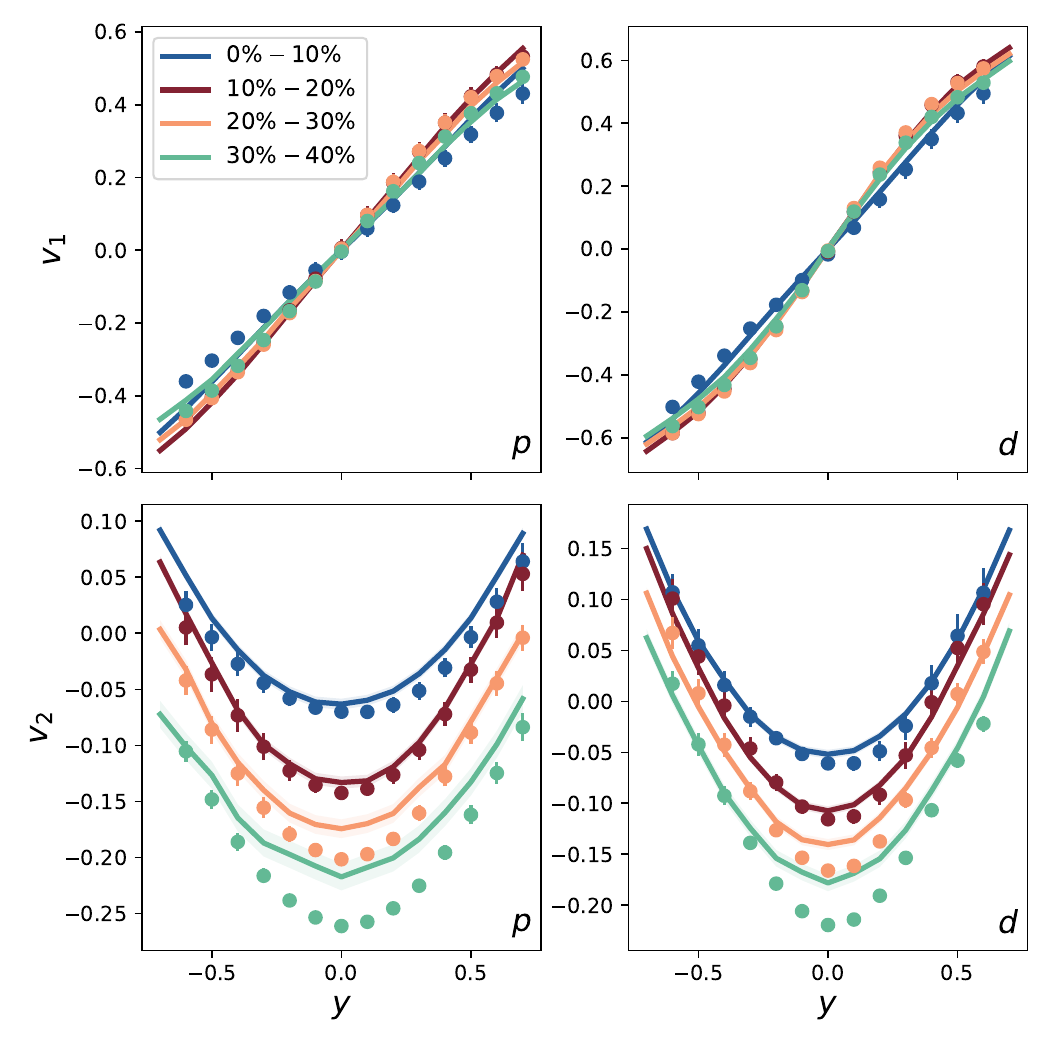}
    \caption{Directed flow $v_1$ and elliptic flow $v_2$ for protons and deuterons as a function of rapidity for $1.0 \,\mathrm{GeV}< p_T < 1.5\,\mathrm{GeV}$ in gold-gold collisions at $E_\mathrm{kin}=1.23A \,\mathrm{GeV}$.
    The circles represent experimental data \cite{HADES:2020lob} and the lines a re predictions from the Gaussian process trained on SMASH calculations. The Gaussian process is, for validation, evaluated in the center of the parameter space $\kappa=300.5 \,\mathrm{MeV},\ S_\mathrm{Pot}=15\,\mathrm{MeV}$.}
    \label{fig:observables}
\end{figure}

Let us start this section with a plot of the experimental data \cite{HADES:2020lob} that we include in the Bayesian analysis.
It is shown in the four panels of Figure \ref{fig:observables} together with a prediction from the Gaussian process that was trained on output from the SMASH transport model.
The prediction from the Gaussian process was done for a parameter set located in the center of the covered parameter space to validate that the Gaussian process can provide realistic results before finding the best parameter set.

\begin{figure}
    \centering
    \includegraphics[width=\columnwidth]{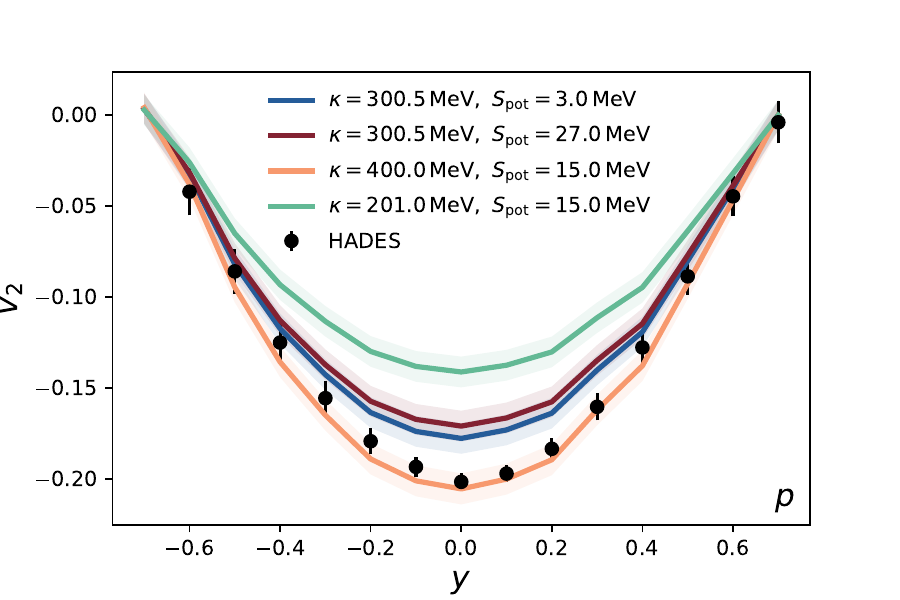}
    \caption{Elliptic flow $v_2$ of protons as a function of rapidity for $1.0 \,\mathrm{GeV}< p_T < 1.5\,\mathrm{GeV}$ in the $20\%-30\%$ most central gold-gold collisions at $E_\mathrm{kin}=1.23A \,\mathrm{GeV}$.
    Experimental data \cite{HADES:2020lob} is shown as black circles and compared predictions from the Gaussian process trained on SMASH calculations for different points in the parameter space.}
    \label{fig:v2_proton}
\end{figure}

\begin{figure}
    \centering
    \includegraphics[width=\columnwidth]{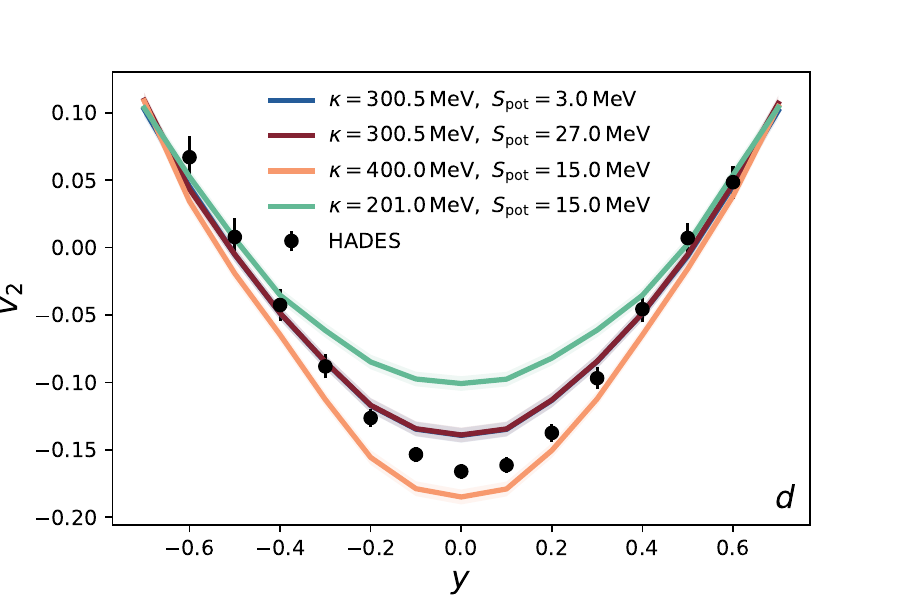}
    \caption{Elliptic flow $v_2$ of deuterons as a function of rapidity for $1.0 \,\mathrm{GeV}< p_T < 1.5\,\mathrm{GeV}$ in the $20\%-30\%$ most central gold-gold collisions at $E_\mathrm{kin}=1.23A \,\mathrm{GeV}$.
    Experimental data \cite{HADES:2020lob} is shown as black circles and compared predictions from the Gaussian process trained on SMASH calculations for different points in the parameter space.}
    \label{fig:v2_deuteron}
\end{figure}

We further test the Gaussian process prediction for different points in the parameter space in Figure \ref{fig:v2_proton}, which shows the elliptic flow of protons in a single centrality class.
Let us start by comparing the curves for the small (red) and large (green) incompressibilities $\kappa$.
We observe that a larger incompressibility leads to a stronger elliptic flow signal which is reassuring because a stiffer equation of state amplifies the squeeze-out effect.
As expected, we observe the same behaviour for deuterons in Figure \ref{fig:v2_deuteron}.
The impact of the symmetry potential is shown Figures \ref{fig:v2_proton} and \ref{fig:v2_deuteron} at a fixed value for the incompressibility.
Comparing the curves for a small (blue) and a large (orange) symmetry potential, we observe no difference for the deuterons.
This is expected as it does not experience any force from that potential since the deuteron does not carry isospin charge.
For protons, we find that the elliptic flow signal is slightly more pronounced with a weaker symmetry potential.
As the gold nuclei have more neutrons than protons, the $I_3$ density is negative in the system.
That means the symmetry potential is attractive for protons, which leads to the smaller elliptic flow that we observe in Fig. \ref{fig:v2_proton}.

\begin{figure}
    \centering
    \includegraphics[width=\columnwidth]{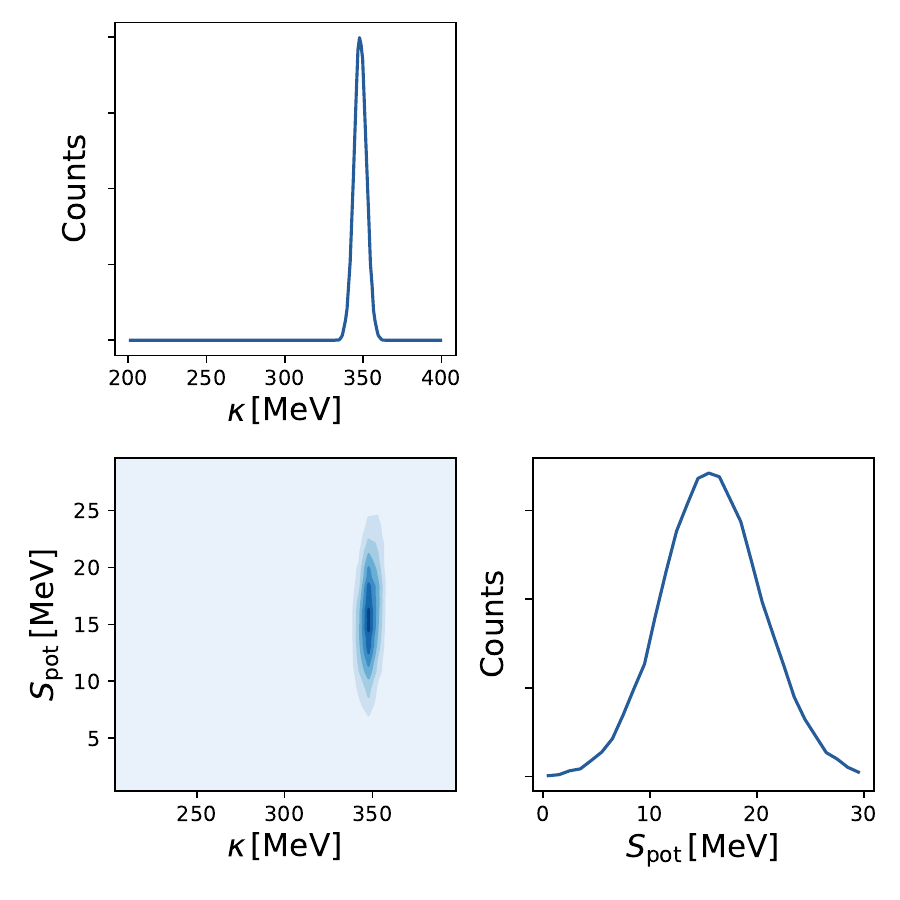}
    \caption{Posterior distribution for the parameters $\kappa$ and $S_\mathrm{Pot}$.
    The bottom left panel shows the posterior distribution as a function of both parameters and the upper and the right panel show the posterior as a function of only $\kappa$ and $S_\mathrm{Pot}$ respectively with the other dependence integrated out.}
    \label{fig:corner}
\end{figure}

As the Gaussian process emulator reproduces the behaviour that we would expect from the transport model, we can perform a Bayesian analysis based on it.
The posterior distribution that we obtain as a result is shown in Figure \ref{fig:corner}.
The $68.3\%$ credibility interval of our parameters is $\kappa = 348.2 ^{+4.0}_{-3.9} \,\mathrm{MeV}$ and $S_\mathrm{Pot} = 15.8^{+4.7}_{-4.6} \,\mathrm{MeV}$.
First thing to note is that we find a relatively stiff equation of state compared to \cite{Danielewicz:2002pu} for example.
According to this analysis, there is only a small uncertainty on the value for the incompressibility and our results are therefore not consistent with previous results.
We would like to highlight here that the parameter estimation is based on the assumption that the model is flawless which is of course not the case.
The problem that a different equation of state is required to describe the data depending on the applied transport model persists but efforts within the Transport Model Evaluation Project (TMEP) aim to resolve this issue in the future.

The symmetry potential we find is consistent with the default value $S_\mathrm{Pot}=18\,\mathrm{MeV}$ used in previous studies \cite{Mohs:2020awg,Tarasovicova:2024isp} but the uncertainty on this parameter is quite large.
One reason for the low sensitivity is that the system has only a small asymmetry as there is no large excess of neutrons in the gold nuclei compared to protons as it is the case in neutron star mergers for example.
Another reason is that the choice of observables, proton and deuteron flow, is done to mainly obtain information on the stiffness but not so much on the symmetry energy.
The value for $S_\mathrm{Pot}$ can be related to the symmetry energy at saturation density $S_0$.
For the $68.3\%$ credibility interval we find $24.2\,\mathrm{MeV} < S_0 < 33.5\,\mathrm{MeV}$.

\begin{figure}
    \centering
    \includegraphics[width=\columnwidth]{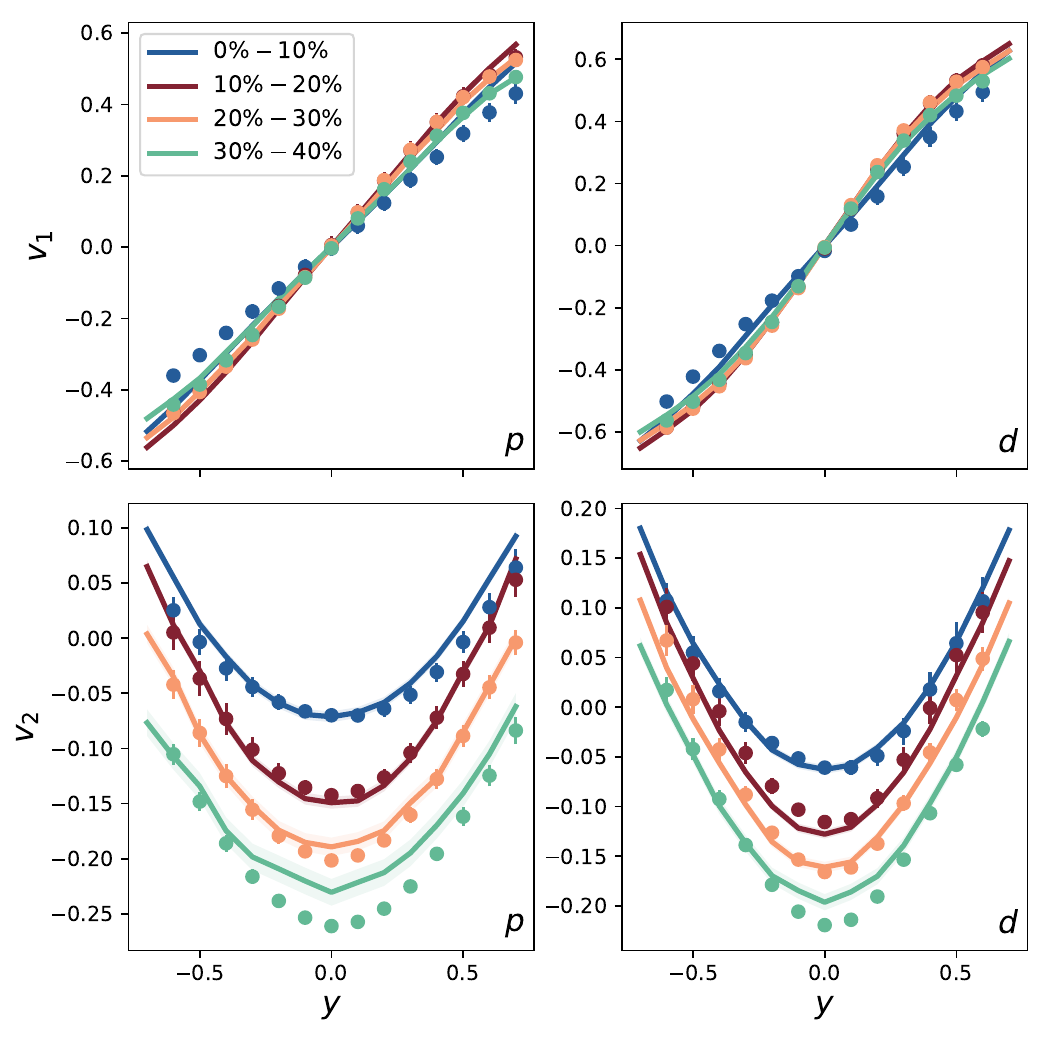}
    \caption{Directed flow $v_1$ and elliptic flow $v_2$ for protons and deuterons as a function of rapidity for $1.0 \,\mathrm{GeV}< p_T < 1.5\,\mathrm{GeV}$ in gold-gold collisions at $E_\mathrm{kin}=1.23A \,\mathrm{GeV}$.
    The circles represent experimental data \cite{HADES:2020lob} and the lines are predictions from the Gaussian process trained on SMASH calculations. The Gaussian process is evaluated at the maximum of the posterior distribution $\kappa=348.2 \,\mathrm{MeV},\ S_\mathrm{Pot}=15.8\,\mathrm{MeV}$.}
    \label{fig:max_posterior}
\end{figure}

We take the maximum of the posterior distribution and calculate the observables based on the Gaussian process as shown in Figure \ref{fig:max_posterior}.
As the prior was chosen to be flat, the maximum of the posterior is the parameter set that has the best agreement with the experimental data.
We find a good overall agreement with the experimental data.
This is on one hand expected as we fit the potentials to match the flow data but on the other hand reassuring that a fair description for so many data points was achieved by varying only two parameters.

\begin{figure}
    \centering
    \includegraphics[width=\linewidth]{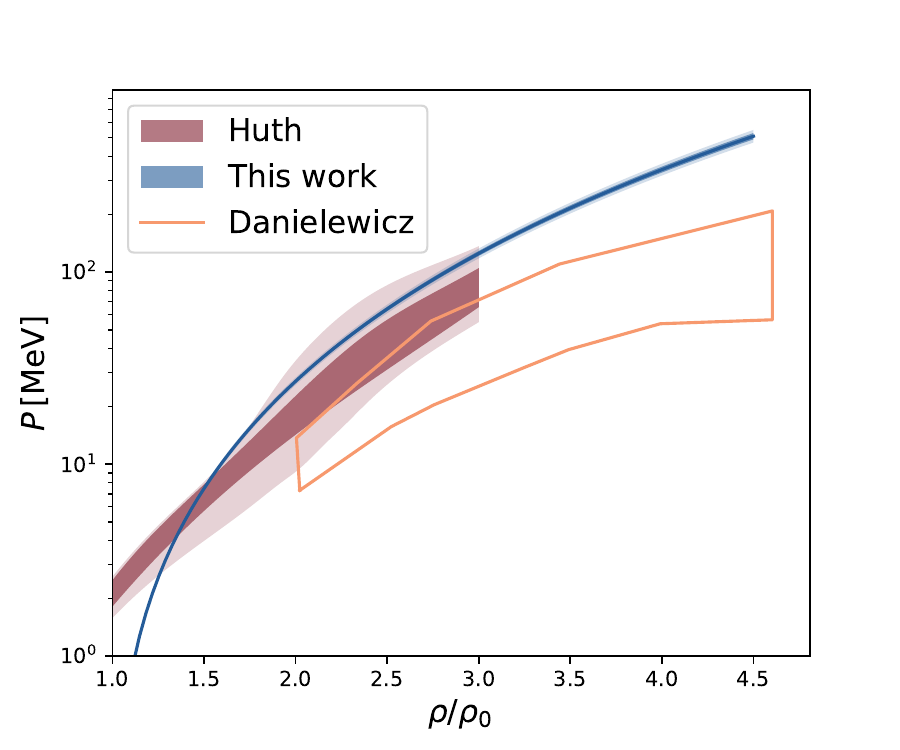}
    \caption{Equation of state of symmetric nuclear matter at vanishing temperature as a function of the baryon density.}
    \label{fig:eos_comparison}
\end{figure}

The equation of state for symmetric nuclear matter at vanishing temperature can be calculated from the estimated parameters.
It is given in terms of the pressure as a function of the baryon density in Figure \ref{fig:eos_comparison} in comparison with the estimate from Danielewicz \cite{Danielewicz:2002pu} and Huth \cite{Huth:2021bsp}.
Especially compared to the work by Danielewicz, we obtain a relatively stiff equation of state.
This may be related to the large amount of resonances present in the SMASH model.
The influence of the resonance content in the calculation on the conclusions regarding the EoS has been reported in \cite{Hombach:1998wr}.
Another difference of the models that can alter the required EoS to describe data is the assumed form of the potentials.
Figure \ref{fig:eos_comparison} contains only information at zero temperature.
Since the nuclear matter is probed in a hot system in heavy-ion collisions, a different form of the potential can lead to a different equation of state at vanishing temperature.

In addition to the difference in the pressure, one can observe that the statistical uncertainty from this work presented in Figure \ref{fig:eos_comparison} is very small.
This is due to the fact that only statistical uncertainties are included in the determination of the posterior distribution but systematic model uncertainties are not taken into account.
The small width of the posterior distribution demonstrates how much information is contained in the experimental data, showing a great perspective for future studies once models have converged in their predictions for flow observables.

\section{Meson Spectra}
\begin{figure*}
    \centering
    \includegraphics[width=0.7\linewidth]{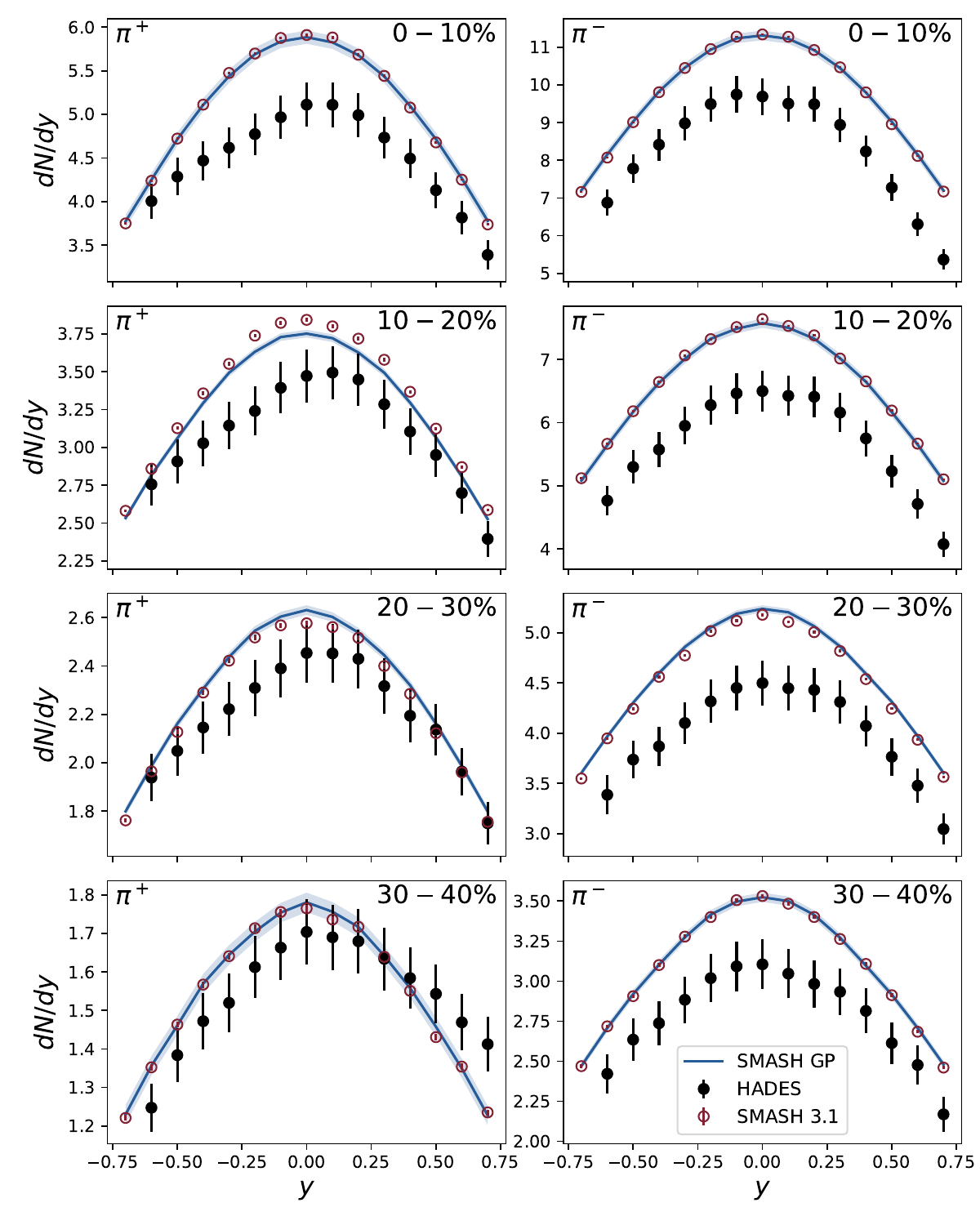}
    \caption{Rapidity spectra for positively and negatively charged pions for different centrality classes compared to experimental data from the HADES experiment \cite{HADES:2020ver}.
    A direct calculation with the SMASH transport model is presented together with the prediction from the Gaussian process that was trained on SMASH output.
    The incompressibility and symmetry potential are set to $\kappa = 349.5 \,\mathrm{MeV}$and $S_\mathrm{pot}=18.16\,\mathrm{MeV}$ respectively.}
    \label{fig:pion_comparison}
\end{figure*}

In this section, we would like to present calculations for the yield of pions and kaons within the improved settings of SMASH including nuclear mean fields compatible with the flow measurements.
While the kaon production is known as an excellent probe for the equation of state, the pion yield is not often considered as a suitable observable.
In the SMASH model, there is a strong sensitivity of the pion yield to the EoS and therefore the number of pions that are obtained when including the momentum-dependent potentials with the parameters that best describe the flow measurement is presented.
The incompressibility and symmetry potential are set to $\kappa = 349.5 \,\mathrm{MeV}$and $S_\mathrm{pot}=18.16\,\mathrm{MeV}$ respectively, which is close to the maximum of the posterior from Section \ref{sec:results}\footnote{The parameters of the potential slightly differ from the MAP values because they were obtained in a previous analysis}.
In Figure \ref{fig:pion_comparison}, the rapidity spectra for positively and negatively charged pions is compared to experimental data in different centrality classes.
For the model, we present the direct calculation using SMASH and a Gaussian process that is trained on the model output and agrees with the direct model calculation.
We choose not to include the pion spectra in the Bayesian analysis due to the observed overshoot in the pion yield compared to the measurement.
The pions in the SMASH model do not experience nuclear potentials and, as there is some tension between the pion multiplicity measured at the HADES and at the FOPI experiment, the pions are not taken into account in the likelihood function. 

\begin{figure}
    \centering
    \includegraphics[width=\linewidth]{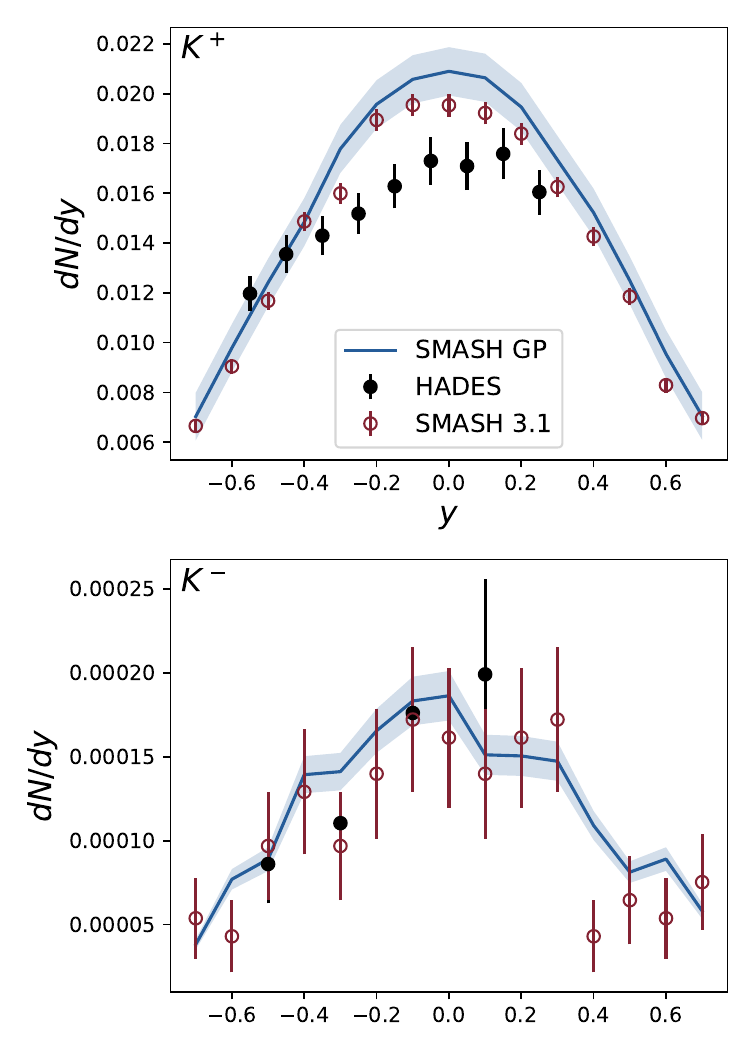}
    \caption{Rapidity spectra of positively and negatively charged kaons in the $0-40\%$ centrality interval compared to experimental data \cite{HADES:2017jgz}.
    A direct calculation with the SMASH transport model is presented together with the prediction from the Gaussian process that was trained on SMASH output.
    The incompressibility and symmetry potential are set to $\kappa = 349.5 \,\mathrm{MeV}$and $S_\mathrm{pot}=18.16\,\mathrm{MeV}$ respectively 
    }
    \label{fig:kaon_comarison}
\end{figure}

In the current version of the SMASH model, kaons also do not experience potentials.
Strangeness potentials are considered to be an important ingredient for describing the production of kaons.
A strong influence of the inclusion of the momentum-dependent term as well as variation of the stiffness of the equation of state on the kaon yield is observed in the calculations.
As similar dependence was obtained in a QMD calculation in \cite{Hartnack:1993bq}.
Without strangeness potentials, the kaons are not included in the Bayesian analysis but the yield is presented in Figure \ref{fig:kaon_comarison}.

The positively charged kaons slightly overshoot the experimental measurement but a reasonable number of kaons is found.
The rapidity distribution for the positively charged kaons is found to be more tight than observed in the experiment.

The multiplicity of negatively charged kaons is very small.
One can see that the kaon yield is quite well described and the shape of the spectrum seems to be described better then for the positively charged kaons.
The limited amount of statistics does however not allow for more definite statements.
In future work, including the kaon yield is certainly interesting as it provides a different angle to study the equation of state in heavy-ion collisions.
Even though the yield can be described consistently with the flow measurements, one should investigate the influence of strangeness potentials before including them in the analysis.

\section{Summary and Conclusions}
\label{sec:conclusions}
We applied the SMASH transport approach to obtain constraints on the equation of state from flow measurements at the HADES experiment.
The transport model includes a Skyrme potential with a momentum-dependent part and a simple term for the symmetry potential.
Light nuclei formation is performed via a coalescence approach and we compare the model to directed and elliptic flow for protons and deuterons in gold-gold collisions for multiple centrality classes.
Constraints were obtained for the incompressibility of nuclear matter at saturation density and the symmetry potential.
Performing a Bayesian analysis, we find a tight constraint for a relatively stiff equation of state and a more loose constraint for the symmetry potential.

Our findings are based on one approach with all its limitations. Therefore, it would be interesting to extend such a Bayesian analysis to several transport approaches and report a global systematic uncertainty in the future. Another aspect left for future work is the exploration of how much additional experimental data at other energies can tighten the constraints. 

\section*{Acknowledgements}
HE, JM and SS acknowledge the support by the State of Hesse within the Research Cluster ELEMENTS (Project ID 500/10.006). Discussions with J. Stroth, T. Galatyuk, M. Lorenz and B. Kardan are greatly acknowledged. 
Computational resources have been provided by the GreenCube at GSI.

\bibliography{bibliography}

\end{document}